# Delicious ice Creams in Plain Awful
## Why does salt thaw ice?

Franco Bagnoli,
Dept. of Physics and Astronomy and Center for the Study of Complex Dynamics, University of Florence, Italy
Via G. Sansone, 1 50019 Sesto Fiorentino (FI) Italy
franco.bagnoli@unifi.it

In a previous issue [1], we exploited a comic by Carl Barks, "The Big Bin on Killmotor Hill", as a starting point for illustrating the anomaly of water, i.e., that water expands when freezing. We can continue here with Barks' most passionate fan, Keno Don Rosa[1], examining his "Return to Plain Awful" [1], which is a sequel of Barks' "Lost in the Andes!"[2], conceived as a tribute for the 40th anniversary of this comics.

Plain Awful is an imaginary valley on the Andes populated by a highly-imitative, cubical people for which the most criminal offence is to exhibit round objects. The duck family (Scrooge, Donald and nephews) are teaming against Scrooge's worst enemy, Flintheart Glomgold, trying to buy the famous Plain Awful square eggs. Inadvertently, Scrooge violates the taboo, showing his Number One Dime, and is imprisoned in the stone quarries. He can be released only after the presentation of an ice cream soda to the President of Plain Awful. Donald and his nephews fly with Flintheart to deliver it, but Scrooge's enemy, of course, betrays the previous agreement after getting the ice cream, forcing the ducks into making an emergence replacement on the spot. Using dried milk, sugar and chocolate from their ration packs, plus some snow and salt for cooling they are able make the ice cream, and after dressing it with the carbonated water from a fire extinguisher they finally manage to produce the desired dessert.

This comic may serve as an introduction to the "mysterious" phenomenon that added salt melts the ice and, even more surprising, does it by lowering the temperature of the mixture. This cooling that has been exploited since a long time for producing homemade ice cream [3], and for self-injuring [4]. Clearly, for a physics show, publicly producing almost-instant ice cream with kitchen equipment (i.e., without liquid nitrogen) is an easy way of getting attendees' attention.

This phenomenon is not so easy to explain because one cannot simply rely on energy considerations. Indeed, water molecules and salt ions (sodium and chlorine) prefer to stay separate, from an energy point of view and so they do, below -21 degrees.[2]

We can illustrate the problem using the same Mercedes-Benz model [6] that we also used for illustrating the water anomaly [1].

---

[1] Don Rosa used to insert the "D.U.C.K" acronym (Dedicated to Uncle Carl by Keno) in the first panel of most of his stories.

[2] Therefore, it is useless to spread salt on street ice below such temperature.

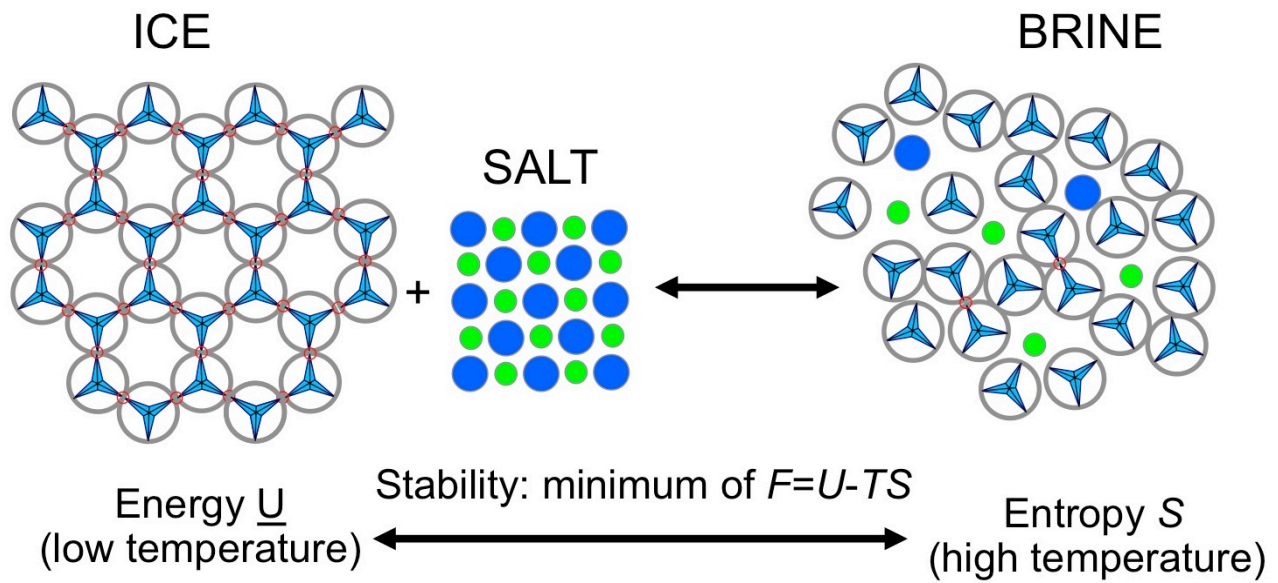

Figure 1: A schematic representation of the mixture of water and salt ions using the Mercedes-Benz model.

As can be seen in Figure 1, the chlorine and sodium ions fit well (being charged) among water molecules, that are polar, but so doing they disrupt the ordered structure of ice and salt crystals. As for melting, this implies that the mixture is stable at high temperatures, and that the crystals are stable at low temperatures. So why does the temperature lowers when salt is mixed with ice?

Let us come back to the previous statement: disordered structures are stable at high temperatures, and ordered ones at low temperatures. This latter is clearly favoured by energy, but the first? We have to introduce the concept of entropy $S$, which is the (logarithm of) the number of possible configurations.

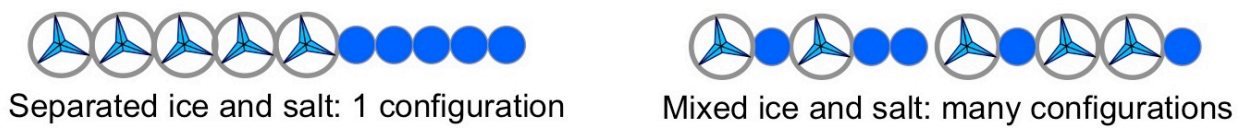

Separated ice and salt: 1 configuration       Mixed ice and salt: many configurations

Figure 2: Number of configuration in ordered and disordered arrangements of molecules and ions in a one-dimensional model.

Using a simple one-dimensional model (Fig. 2), we can appreciate the difference in the number of configurations for ordered and disordered structures. If the difference in energy between the two types of structures is large, the order is preferred with occasional and local fluctuations (for temperatures above the absolute zero and in a classical framework). But if the difference in energy is not so large, fluctuations that rise energy happen more frequently and when order is destroyed, it is recovered with difficulty because there are so many un-ordered configurations nearby, not so distant in energy, and only one ordered.

Scientists use to introduce the concept of free energy $F = U - TS$, where $U$ is the energy and $T$ the temperature. The stability of a system is given by the minimum of $F$, and from the above formula it is easy to see that the temperature favours the energy when it is low, and the entropy (that has a minus sign) when it is large.

So, adding salt to ice above -21 degrees, molecules tend to reach a stable configuration by mixing, but this requires energy in order to break the hydrogen bonds of ice and for the dissolution of the salt (the latent heat of ice is 6.01 kJ/mol and the dissolution of sodium chloride requires 3.87 kJ/mol).

And now our challenge (for people using the International System of Units): why is the Fahrenheit scale so weird and why does its zero correspond to (roughly) -18 degrees?[3]

---

[3] At the time of Daniel Gabriel Fahrenheit (1686–1736), the mixture of salt, ice and water constituted the lowest temperature easily obtainable without laboratory tools, and was used to define the zero of the Fahrenheit scale (using equal volumes of ice and salt), while 100 degrees corresponded (roughly) to the body temperature [8]. The advantages of this scale is that it never requires negative temperatures in everyday life, and (with the present refined scale) that there is a 180 degree separation between the boiling and freezing points of water, making easier to measure temperatures without decimals.